# Determination of the thermopower of microscale samples with an AC method

Hanfu Wang[1,a)], Fanglong Yang[1,2], Yanjun Guo[1], Kaiwu Peng,[1] Dongwei Wang[1], Weiguo Chu[1,b)] Shuqi Zheng[2,c)]

[1] CAS Key Laboratory of Nanosystem and Hierarchical Fabrication，CAS Center of Excellence for Nanoscience，National Center for Nanoscience and Technology, Beijing 100190, P. R. China.

[2] Department of Materials Science & Engineering, College of Science, China University of Petroleum-Beijing, No. 18 Fuxue Road, Changping, Beijing 102249, China

a)Electronic mails: wanghf@nanoctr.cn ;  b) Electronic mail: wgchu@nanoctr.cn ;  c) Electronic mail: zhengsq09@163.com.

**Abstract:** A modified AC method based on micro-fabricated heater and resistive thermometers has been applied to measure the thermopower of microscale samples. A sinusoidal current with frequency ω is passed to the heater to generate an oscillatory temperature difference across the sample at a frequency 2ω, which simultaneously induces an AC thermoelectric voltage, also at the frequency 2ω. A key step of the method is to extract amplitude and phase of the oscillatory temperature difference by probing the AC temperature variation at each individual thermometer. The sign of the thermopower is determined by examining the phase difference between the oscillatory temperature difference and the AC thermoelectric voltage. The technique has been compared with the popular DC method by testing both n-type and p-type thin film samples. Both methods yielded consistent results, which verified the reliability of the newly proposed AC method.

## I. INTRODUCTION

The thermopower, also known as Seebeck coefficient, serves not only as a performance indicator of thermoelectric materials, but also as a sensitive probe to reveal transport properties and electronic structures of semiconductors and metals around the Fermi levels. Measurement of thermopower typically involves generating a small temperature gradient across the sample, and recording the temperature difference as well as thermally induced thermoelectric voltage. During the measurement, it is crucial to reliably obtain the temperature difference. For bulk samples, the temperature measurements are routinely accomplished by making two pairs of thermocouples in close contact with the sample. [1-9] The thermopower measurements can then be carried out in steady-state (DC)[1, 3], quasi-steady state [2, 5-6] or AC configurations[7-9]. [4] As the size of the sample shrinks to microscale or nanoscale, two four-probe resistive thermometers are often micro-fabricated to sense the temperatures at two ends of the sample which is typically loaded on a substrate. In this case, the DC method is the most popular way to obtain the thermopower. [10-16] By sourcing a DC current to a micro-heater located in proximity to one end of the sample, a stable temperature difference $\Delta T_{DC}$ is established along the sample and the corresponding thermoelectric voltage $\Delta U_{DC}$ can be measured by a voltmeter. The thermopower S is extracted from the slope of a best-fit line of $\Delta U_{DC}$ as a function of $\Delta T_{DC}$ in order to remove possible voltage offsets arising from the measurement circuit. This DC method is somehow time-consuming since multiple pairs of $\Delta U_{DC} \sim \Delta T_{DC}$ data need to be recorded. The problem may become more serious if the thermopower is scanned as a function of external fields (e.g. gate voltage[10]).

Alternatively, an AC measurement method based on the same configuration of one heater and two four-probe thermometers has been proposed in the literature. [17-19] In this method, an AC current of frequency $\omega$ is supplied into the heater to create an oscillatory temperature difference along the sample at a frequency $2\omega$. The resulted $2\omega$ thermoelectric voltage $\Delta U(2\omega)$ is detected by a lock-in amplifier. In Ref. 18, the thermopower $S_{AC}$ was given in the form of

$$S_{AC} = -\Delta U(2\omega)/\Delta T(2\omega), \qquad (1)$$

where $\Delta T(2\omega)$ is the temperature difference. It should be noted that, under the AC heating, the temperature difference includes not only the AC component, but also a DC component. As a result, the induced thermoelectric voltage consists of both AC and DC components. Since the AC component of the thermoelectric voltage appears on the numerator of Eq. (1), the corresponding AC component of the temperature difference should be the applied to the denominator. In Ref.18, $\Delta T(2\omega)$ was determined from $\Delta T(2\omega) = T_H(2\omega) - T_L(2\omega)$, where $T_H(2\omega)$ ($T_L(2\omega)$) is supposed to be the amplitude of the temperature oscillation for the thermometer at the hot (cold) end. As mentioned later in Section II, the amplitude of the AC temperature difference actually depends not only on the amplitude but also on the phase of the temperature oscillation of each thermometer.

In this paper, we present a modified AC method to measure the thermopower of the microscale samples by explicitly retrieving the amplitude and phase of the 2ω temperature difference. To achieve this, the amplitude and phase of the AC temperature oscillation are measured for each individual thermometer. The results are used to yield the amplitude and phase of the 2ω temperature difference through a simple trigonometric function calculation. Also, the amplitude and phase of the 2ω thermoelectric voltage are directly recorded by the lock-in amplifier. The amplitude and the sign of the resulted thermopower depend on the amplitude ratio and phase relationship between the 2ω thermoelectric voltage and the 2ω temperature difference, respectively. The method, along with the popular DC method, has been applied to the p-type $Sb_2Te_3$ and n-type $Bi_2Te_3$ thin film test samples. The results of the two methods agree with each other reasonably well within experimental uncertainty.

## II. EXPERIMENTAL METHODOLOGY

As illustrated in Fig. 1, when a sinusoidal current with an angular frequency ω is driven through the heater, a periodic heating wave at the frequency 2ω will be generated via Joule heating and it will propagate towards the microscale sample through the substrate. The sample is assumed to be in good thermal contact with the substrate and with the two thermometers Th1 and Th2, of which Th1 is located closer to the heater. To sense the temperature variation, two outer contact pads of Th1 or Th2 are used for sourcing current, and two inner contact pads for measuring resistive voltage drop.

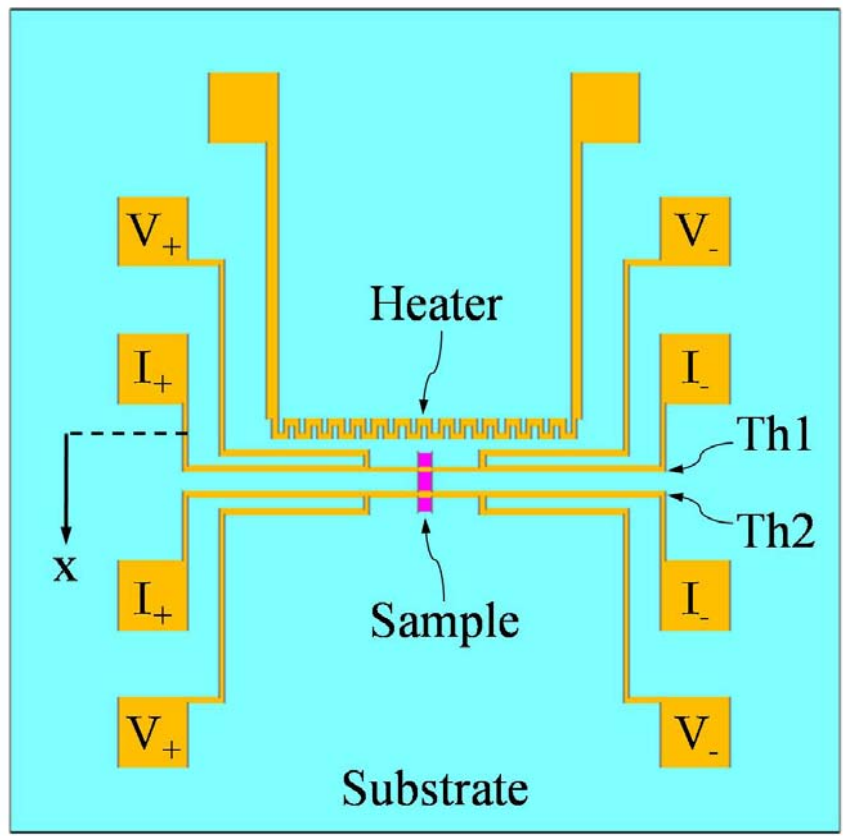

FIG. 1  Layout of the device for measuring the thermopower of the microscale sample.

The instantaneous temperatures at the positions where the two thermometers are located can be expressed as

$$T_i = T_0 + \Delta T_i^{dc} + T_i^{ac} = T_0 + \Delta T_i^{dc} + |T_i^{ac}| \cdot \sin(2\omega t + \varphi_i) \quad (i=1, 2), \quad (2)$$

where subscripts 1 and 2 correspond to Th1 and Th2, respectively. $T_0$ is the initial temperature of the substrate before turning on the heater. $\Delta T_i^{dc}$ represents the DC temperature rise of the thermometer, $T_i^{ac}$ the AC temperature fluctuation with the angular frequency $2\omega$. $|T_i^{ac}|$ and $\varphi_i$ are the amplitude and phase of $T_i^{ac}$, respectively.

If we approximate the heat transfer along x-direction through the substrate surface as a one-dimensional (1D) process, the AC component of the temperature variation at position x (the heater is assumed to be located at x = 0) can be given as [20-21]

$$T_{1D}^{ac} = \frac{j_0}{k}\sqrt{\frac{\alpha}{2\omega}} \exp\left(-\sqrt{\frac{\omega}{\alpha}}x\right)\cos\left(2\omega t - \sqrt{\frac{\omega}{\alpha}}x - \frac{\pi}{4}\right), \quad (3)$$

where $j_0$ is heat flux amplitude. k and α are the thermal conductivity and the thermal diffusivity of the substrate, respectively. Eq. (3) suggests that the thermal wave propagates away from the heater with a decaying amplitude and a phase shift, which further implies that $T_1^{ac}$ and $T_2^{ac}$ in Eq. (2) should have not only different amplitude ($|T_1^{ac}| > |T_2^{ac}|$), but also different phase ($\varphi_1 \neq \varphi_2$). The scenario is schematically shown in Fig. 2(a) and Fig. 2(b).

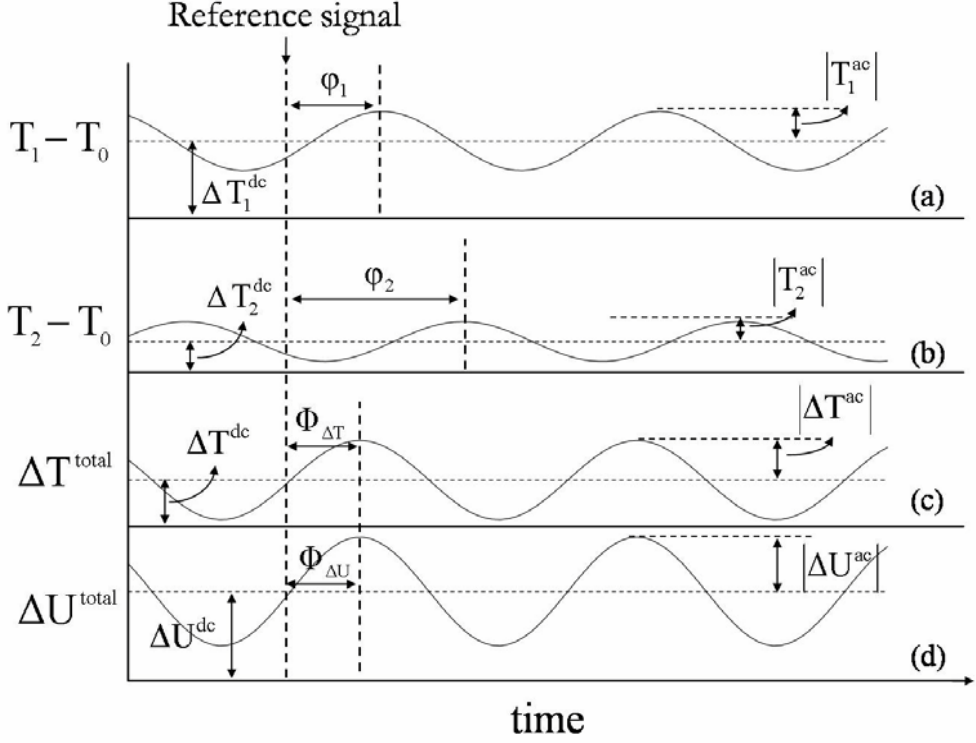

FIG. 2 Schematic illustration of instantaneous temperature of Th1(a) and Th2(b)、temperature difference $\Delta T^{total}$(c) and thermoelectric voltage $\Delta U^{total}$ (in the case of a n-type sample) (d) as a function of time.

Starting from Eq. (2), one may obtain the temperature difference $\Delta T^{total}$ between T1 and T2 by performing a simple trigonometric function calculation. The result can be expressed as

$$\Delta T^{total} = \Delta T^{dc} + \Delta T^{ac} = \Delta T^{dc} + |\Delta T^{ac}| \cdot \sin(2\omega t + \Phi_{\Delta T}), \qquad (4a)$$

where

$$|\Delta T^{ac}| = \sqrt{(|T_1^{ac}| \cdot \cos\varphi_1 - |T_2^{ac}| \cdot \cos\varphi_2)^2 + (|T_1^{ac}| \cdot \sin\varphi_1 - |T_2^{ac}| \cdot \sin\varphi_2)^2}, \qquad (4b)$$

$$\cos\Phi_{\Delta T} = (|T_1^{ac}| \cdot \cos\varphi_1 - |T_2^{ac}| \cdot \cos\varphi_2)/|\Delta T^{ac}|, \qquad (4c)$$

$$\sin\Phi_{\Delta T} = (|T_1^{ac}| \cdot \sin\varphi_1 - |T_2^{ac}| \cdot \sin\varphi_2)/|\Delta T^{ac}|. \qquad (4d)$$

Here, $\Delta T^{dc}$ and $\Delta T^{ac}$ are the DC and AC component of the temperature difference, respectively. $|\Delta T^{ac}|$ and $\Phi_{\Delta T}$ are the amplitude and phase of

$\Delta T^{ac}$, respectively. Eq. (4b) suggests that the amplitude of the AC temperature difference is determined by both amplitude and phase of the $2\omega$ temperature oscillations of the two thermometers.

Due to the existence of $\Delta T^{total}$, the induced thermoelectric voltage $\Delta U^{total}$ can be expressed as

$$\Delta U^{total} = \Delta U^{dc} + \Delta U^{ac} = \Delta U^{dc} + |\Delta U^{ac}| \cdot \sin(2\omega t + \Phi_{\Delta U}), \quad (5)$$

where $\Delta U^{dc}$ and $\Delta U^{ac}$ are the DC and AC component of the thermoelectric voltage, respectively. $|\Delta U^{ac}|$ and $\Phi_{\Delta U}$ are the amplitude and phase of $\Delta U^{ac}$, respectively.

Finally, the Seebeck coefficient $S_{AC}$ can be found from

$$S_{AC} = \pm |\Delta U^{ac}| / |\Delta T^{ac}|, \quad (6)$$

where the sign of $S_{AC}$ is determined from the phase relationship between $\Delta U^{ac}$ and $\Delta T^{ac}$. For a n-type sample, $\Delta U^{ac}$ and $\Delta T^{ac}$ vary with time in phase and $|\Phi_{\Delta U} - \Phi_{\Delta T}|$ is equal to 0°, as shown in Fig. 2(c) and Fig. 2(d). Therefore, $S_{AC}$ will be endowed with a negative sign. In the case of a p-type sample, $\Delta U^{ac}$ and $\Delta T^{ac}$ are 180° out-of-phase, which brings a positive sign to $S_{AC}$.

## III. EXPERIMENTAL DETAILS

The experimental setup for carrying out the measurement is schematically shown in Fig. 3. It consists of a SR850 lock-in amplifier, a Keithley 6221 AC/DC current source, a Keithley 2400 source meter and a matrix switching card. The data acquisition is made by a PC computer installed with a Labview software. The Keithley 6221 current source is used for feeding the AC current into the heater.

The major steps of the measurement include obtaining the 2ω temperature fluctuations at the two thermometers, i.e., $T_1^{ac}$ and $T_2^{ac}$, and the 2ω thermoelectric voltage $\Delta U^{ac}$ in sequence. The matrix switching card is used to connect the selected contact pads of the thermometers to the proper instruments.

For example, in order to measure $T_1^{ac}$, the matrix switching card is configured in such a way that the two outer contact pads (P2 and P4) of Th1 are connected to the output terminals of the Keithley 2400 source meter, while the two inner contact pads (P1 and P3) are linked to the differential input terminals of the SR850 lock-in amplifier. The contact pads of Th2 are not connected to any instruments. The Keithley 2400 source meter passes a DC "probe" current $I_0$ through Th1. Since the AC temperature component sensed by the thermometer oscillates at the frequency 2ω and the thermometer has a non-zero temperature coefficient of resistance, the resistance of the thermometer gains a small AC component $R_1^{ac}$ oscillating at the frequency 2ω. As a result, a small AC voltage at the frequency 2ω arises by multiplication of the AC resistance with the "probe" current. [22] The amplitude $\left|V_1^{ac}\right|$ and phase $\varphi_1$ of the AC voltage are detected by the SR850 lock-in amplifier. During the measurement, the phase $\varphi_1$ is referred to an external reference signal provided by the Keithley 6221 current source which is synchronized with the lock-in amplifier. Note that $\varphi_1$ is also the phase of the 2ω temperature fluctuation of Th1.

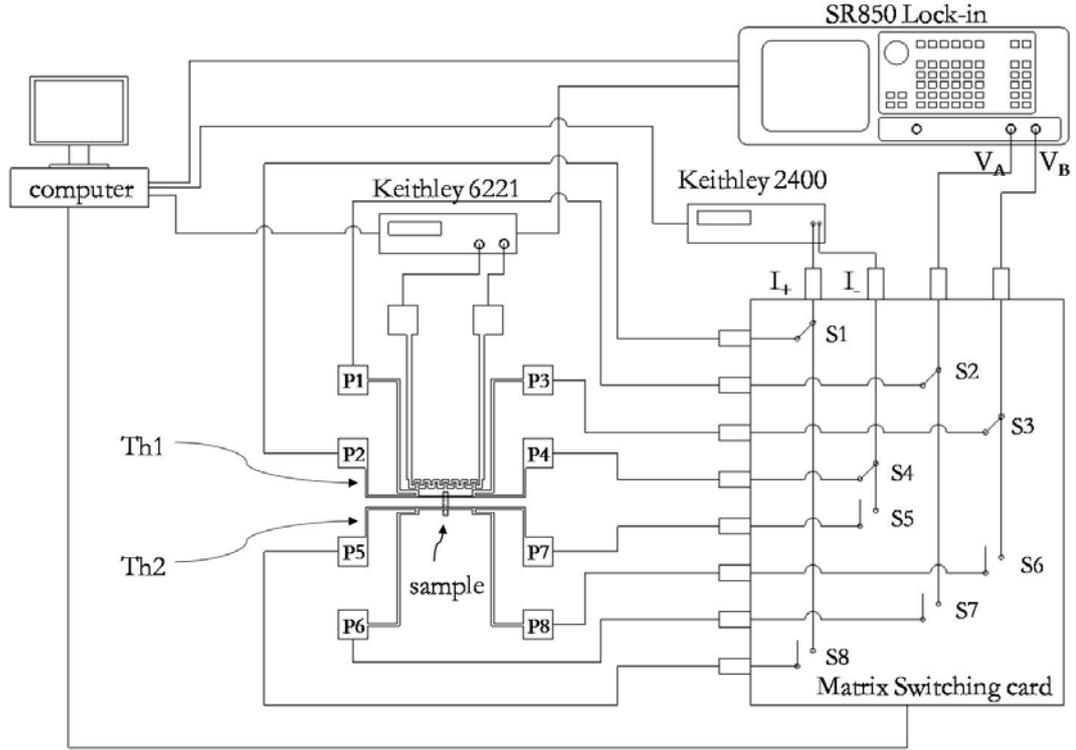

FIG. 3 Schematic of the experimental setup for measuring the thermopower of the microscale sample.

The amplitude of the 2ω temperature fluctuation at Th1 is calculated by

$$\left|T_1^{ac}\right| = \left|R_1^{ac}\right|/(dR_1/dT) \quad (7a)$$

and

$$\left|R_1^{ac}\right| = \left|V_1^{ac}\right|/I_0, \quad (7b)$$

where $dR_1/dT$ is the first derivative of the resistance of the thermometer over temperature for Th1 which is derived from the calibration curve of Th1. $|R_1^{ac}|$ is the amplitude of the oscillatory resistance of Th1. Similarly, $T_2^{ac}$ of Th2 can be obtained by using the same procedure. With $T_1^{ac}$ and $T_2^{ac}$ in hand, $|\Delta T^{ac}|$ and $\Phi_{\Delta T}$ can be computed by using Eqs. (4b)-(4d).

To measure $\Delta U^{ac}$, the contact pads S2 and S6 are connected to the differential input terminals of the SR850 lock-in amplifier and all other contact pads of Th1 and Th2 are left unconnected. The amplitude $|\Delta U^{ac}|$ and the phase $\Phi_{\Delta U}$ are recorded directly by the lock-in amplifier and $\Phi_{\Delta U}$ is also referred to the reference signal generated by the Keithley 6221 current source. It should be mentioned that since the SR-850 lock-in amplifier is designed to measure the root-mean-squared (RMS) values of the voltage signals, the measured signals (e. g., $|\Delta U^{ac}|$) and the values derived from the measurements (e. g., $|\Delta T^{ac}|$) are all expressed in RMS units in this paper.

With the experimental setup described above, we have measured the thermopowers of p-type $Sb_2Te_3$ and n-type $Bi_2Te_3$ thin film samples with a rectangular shape (about 100 μm in width and 800 μm in length). The samples were prepared by thermal evaporation of high purity $Sb_2Te_3$ or $Bi_2Te_3$ powder through a shadow mask onto a glass substrate at room temperature in a high vacuum chamber. Both samples were subsequently annealed at 200 °C at a pressure below $3.8 \times 10^{-3}$ Pa for one hour. The thicknesses of the $Sb_2Te_3$ film and the $Bi_2Te_3$ film are 400 nm and 360 nm, respectively.

To build a test device, another shadow mask on which the patterns of the thermometers and the heater are defined was placed on top of the glass substrate deposited with the film. A home–made shadow mask aligner was employed to align the patterns properly with the rectangular-shaped thin film. The microelectrodes were fabricated by evaporating 200 nm gold through stencil apertures onto the substrate. A T-type thermocouple was then attached to the substrate surface with silver paste for calibrating the two thermometers afterwards.

The device was then attached to a copper sample carrier with a thermally conducting grease (Apiezon H). The sample carrier was subsequently mounted on a home-made vacuum chamber for the thermopower measurements as well as the thermometer calibrations.

## IV. RESULTS

Figure 4 displays an optical microscopy photograph of a typical device used for measuring the thermopower of the thin film sample.

To calibrate the two thermometers (Th1 and Th2) of the device, their resistances were recorded as a function of the substrate temperature measured by the attached T-type thermocouple. Figure 5 illustrates typical calibration curves obtained around the room temperature which demonstrate good linearity in the range of 302K-325K.

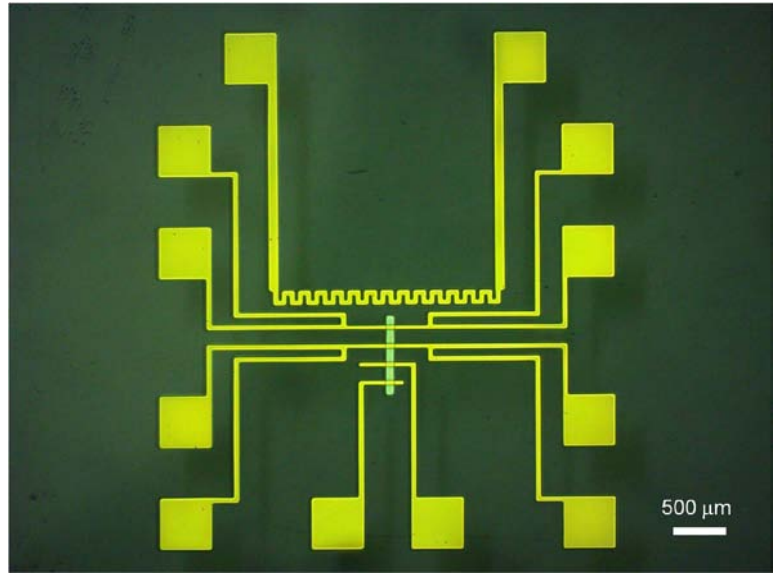

FIG. 4 Optical microscopy image of a representative device for measuring the thermopower of the thin film sample.

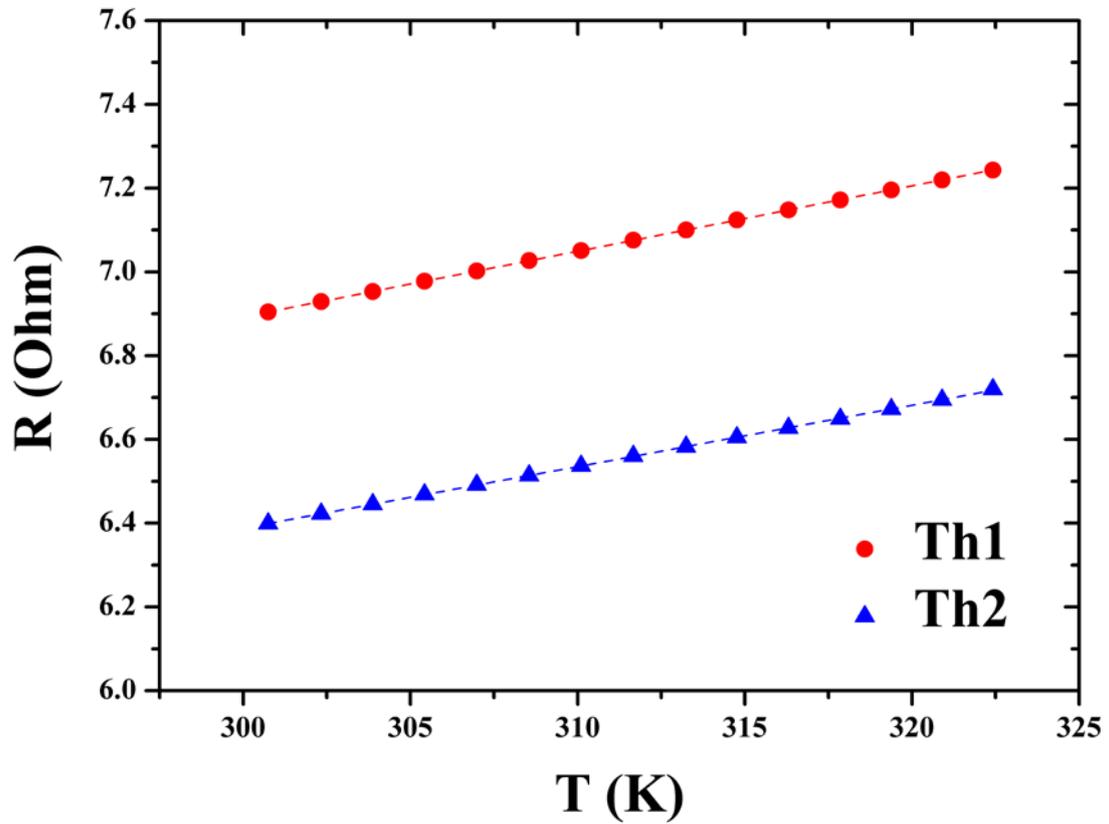

FIG. 5 Typical calibration curves for the thermometers in the device. The dashed lines represent linear best fits to the experimental data.

Figure 6 illustrates the amplitudes of the 2ω temperature oscillations of the two thermometers, namely $|T_1^{ac}|$ and $|T_2^{ac}|$, as a function of frequency of the heating current $f_{HC}$. As the frequency increases, the amplitude decreases quickly. The trend is dominated by the thermophysical properties of the glass substrate, as suggested by Eq. (3). When the frequency is above 3.13 Hz, $|T_2^{ac}|$ will be below 0.05 K（RMS）. To ensure accuracy of the temperature measurement, the frequency of the heating current used in this work didn't exceed 3.13 Hz.

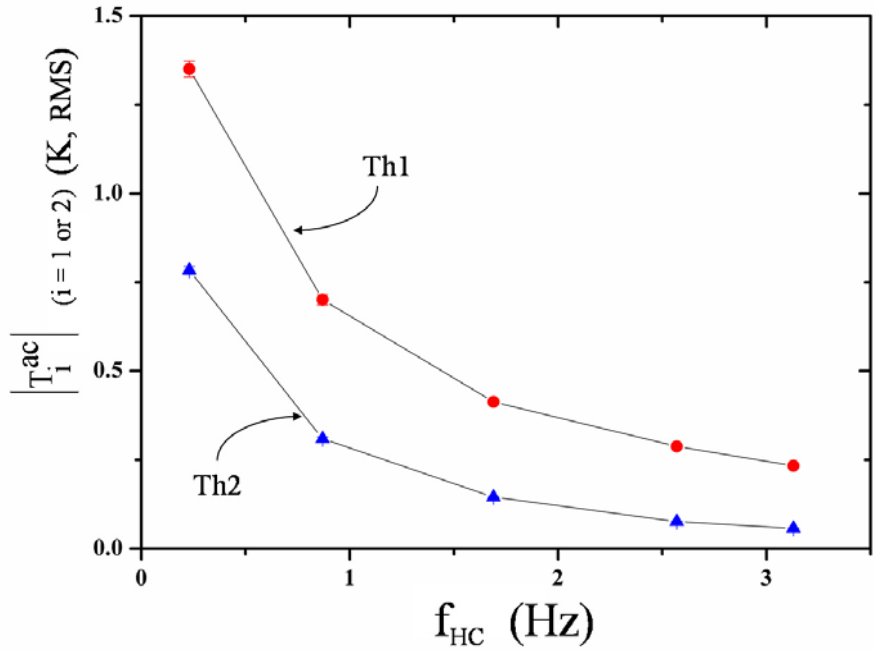

FIG. 6  $|T_1^{ac}|$ and $|T_2^{ac}|$ as a function of frequency of the heating current.

The thermopowers of the $Sb_2Te_3$ and $Bi_2Te_3$ thin films were first measured at $f_{HC}$ = 0.87 Hz around room temperature. Typical intermediate results from a single measurement are summarized in Table 1 for both samples.

Table 1 Typical intermediate results for obtaining the thermopowers of the $Sb_2Te_3$ and $Bi_2Te_3$ films[*]

|  | $Sb_2Te_3$ film | $Bi_2Te_3$ film |
|---|---|---|
| $\left|R_1^{ac}\right|$ (Ohm, RMS) | $1.11 \times 10^{-2}$ | $1.11 \times 10^{-2}$ |
| $(dR_1/dT)^{-1}$ (K/Ohm) | 61.83 | 64.71 |
| $\left|T_1^{ac}\right|$ (K, RMS) | 0.69 | 0.72 |
| $\varphi_1$ (°) | -157.46 | -157.38 |
| $\left|R_2^{ac}\right|$ (Ohm, RMS) | $4.58 \times 10^{-3}$ | $4.63 \times 10^{-3}$ |
| $(dR_2/dT)^{-1}$ (K/Ohm) | 66.27 | 68.86 |
| $\left|T_2^{ac}\right|$ (K, RMS) | 0.30 | 0.32 |
| $\varphi_2$ (°) | 168.60 | 170.04 |
| $\left|\Delta T^{ac}\right|$ (K, RMS) | 0.47 | 0.49 |
| $\Phi_{\Delta T}$ (°) | -136.19 | -136.64 |
| $\left|\Delta U^{ac}\right|$ (μV, RMS) | 51.48 | 53.82 |
| $\Phi_{\Delta U}$ (°) | 44.32 | -137.15 |
| $\left|\Phi_{\Delta U} - \Phi_{\Delta T}\right|$ (°) | 180.51 | 0.51 |

*Heating current: 30mA (RMS), $f_{HC}$ = 0.87 Hz.

From the Table 1, it can be seen that the difference between $\Phi_{\Delta T}$ and $\Phi_{\Delta U}$ is close to 180° for the $Sb_2Te_3$ film, and is close to 0° for the $Bi_2Te_3$ film. The results indicate that the sign of the thermopower should be positive for the $Sb_2Te_3$ film and be negative for the $Bi_2Te_3$ film, as is expected. After multiple measurements, the thermopower at $f_{HC} = 0.87$ Hz was found to be 107.9±2.2 μV/K for the $Sb_2Te_3$ film and −109.5±2.8 μV/K for the $Bi_2Te_3$ film. It should be noted that if the phase lag between the two thermometers was ignored and $|\Delta T^{ac}|$ was obtained directly from the difference of $|T_1^{ac}|$ and $|T_2^{ac}|$, the resulted thermopowers will be overestimated by more than 20% in both cases.

Figure 7(a) and 7(b) reveal the frequency dependence of $|\Delta U^{ac}|$ and $|\Delta T^{ac}|$ for the two samples. Both $|\Delta U^{ac}|$ and $|\Delta T^{ac}|$ decrease with increasing frequency for either sample. The frequency dependence of $S_{AC}$ and $|\Phi_{\Delta U} - \Phi_{\Delta T}|$ is presented in Fig. 7(c), which clearly indicates that both properties are essentially independent of the frequency in the range below 3.13 Hz.

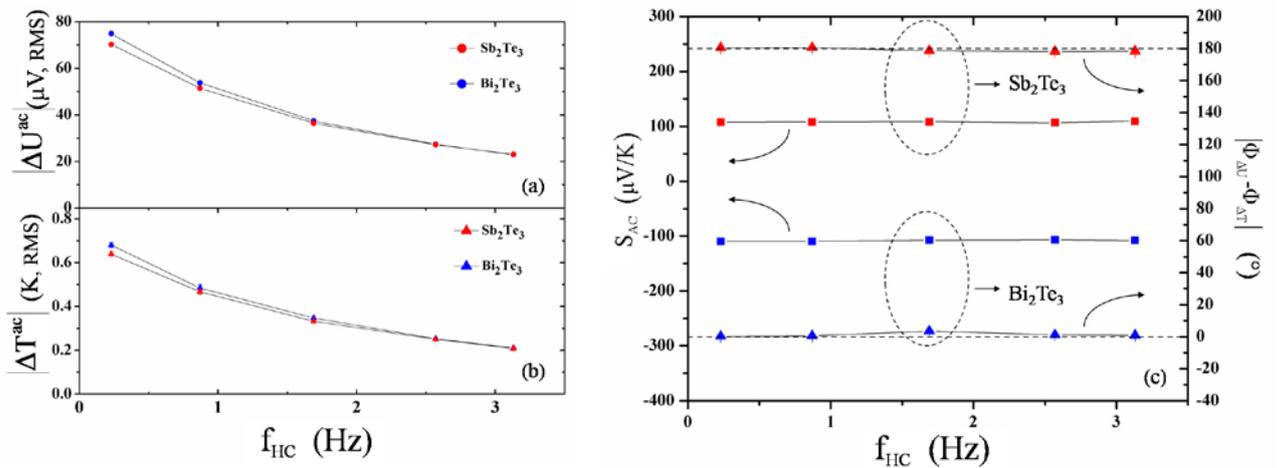

FIG.7 $|\Delta U^{ac}|$ (a)、$|\Delta T^{ac}|$ (b)、$S_{AC}$ and $|\Phi_{\Delta U} - \Phi_{\Delta T}|$ (c) as a function of the frequency of the heating current.

In order to verify the results of the AC measurements, we also carried out the measurements in the DC configuration. In brief, a DC heating current was injected from the Keithley 6221 current source into the heater to produce a stable temperature difference $\Delta T_{DC}$ across the sample. The corresponding thermoelectric voltage $\Delta U_{DC}$ was measured by the Keithley 2400 source meter while $\Delta T_{DC}$ was found from

$$\Delta T_{DC} = \Delta T_{1DC} - \Delta T_{2DC} , \qquad (8)$$

where $\Delta T_{1DC}$ and $\Delta T_{2DC}$ are DC temperature rises of Th1 and Th2, respectively. They were derived from the respective resistance change and calibration curve of Th1 and Th2. The resistances of the thermometers were determined by the standard four-probe method. The SR850 lock-in amplifier sourced a small current into Th1 or Th2 which was electrically connected with a resistor ($R_0 = 1K$ Ohm) in series. The voltage drop $V_{th}$ extracted from the inner pads of the thermometer was measured by the same lock-in amplifier, while the voltage drop $V_R$ across the resistor was measured by another SR830 lock-in amplifier. The resistance of the thermometer $R_{th}$ was calculated from $R_{th} = V_{th} R_0 / V_R$.

Figure 8 demonstrates good linearity of $\Delta U_{DC}$ with respect to $\Delta T_{DC}$ for both $Sb_2Te_3$ and $Bi_2Te_3$ films. From linear best fits to the data, the thermopowers of the $Sb_2Te_3$ and $Bi_2Te_3$ films were found to be 108.6 ± 3.3µV/K and -112.3 ± 1.6 µV/K, respectively. Within measurement uncertainty, the values agree reasonably well with the results obtained by the AC method.

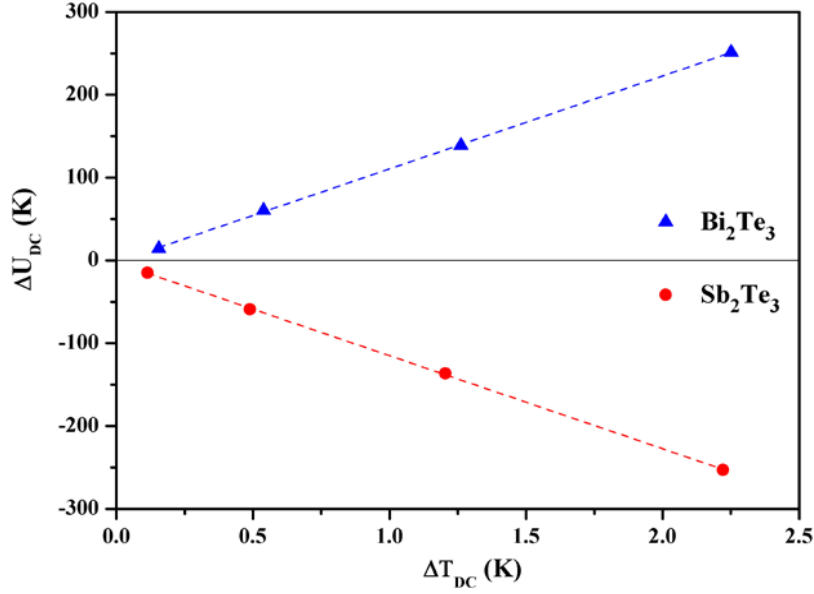

FIG. 8  $\Delta U_{DC}$ as a function of $\Delta T_{DC}$ for the $Sb_2Te_3$ and $Bi_2Te_3$ films. In the measurement, the DC heating current ranges from 10 mA to 40mA. The dashed lines represent linear best fits to the experimental data.

## V. SUMMARY

We have introduced a modified AC method to measure the thermopower of microscale samples. A heating power with frequency $2\omega$ is used to generate the AC temperature difference and the corresponding thermoelectric voltage across the sample. The phase and the amplitude of the thermoelectric voltage can be detected by the lock-in amplifier. The crucial step in this method is the determination of the oscillatory temperature difference by probing the AC temperature variation of each thermometer. The amplitude and the sign of the thermopower depend on the amplitude ratio and phase relationship between the 2ω thermoelectric voltage and the 2ω temperature difference, respectively.

The AC technique has been compared to the well-known DC method. Both methods yield consistent results when measuring the p-type and n-type test samples. The technique introduced in this work may provide a relatively rapid way to probe the thermopowers of the microscale or nanoscale samples, especially when the measurements are undertaken by scanning external electrical field or magnetic field.

**Acknowledgments:**

Wang H. would like to acknowledge the financial support from the National Natural Science Foundation of China under grant No. 61176083, the State Key Development Program for Basic Research of China under grant No. 2011CB932801.